\documentclass[twocolumn,showpacs,preprintnumbers]{revtex4}

\usepackage{graphicx}   
\usepackage{dcolumn}    
\usepackage{bm}         

\begin{document}

\title{Search for photon oscillations into massive particles}

\author{M. Fouch\'e}
\email{mathilde.fouche@irsamc.ups-tlse.fr} \affiliation{
Laboratoire Collisions Agr\'{e}gats R\'{e}activit\'{e}, IRSAMC, UPS/CNRS,
UMR 5589, 31062 Toulouse, France.}
\author{C. Robilliard}
\affiliation{
Laboratoire Collisions Agr\'{e}gats R\'{e}activit\'{e}, IRSAMC, UPS/CNRS,
UMR 5589, 31062 Toulouse, France.}
\author{S. Faure}
\affiliation{
Laboratoire Collisions Agr\'{e}gats R\'{e}activit\'{e}, IRSAMC, UPS/CNRS,
UMR 5589, 31062 Toulouse, France.}
\author{C. Rizzo}
\affiliation{
Laboratoire Collisions Agr\'{e}gats R\'{e}activit\'{e}, IRSAMC, UPS/CNRS,
UMR 5589, 31062 Toulouse, France.}

\author{J. Mauchain}
\affiliation{Laboratoire National des Champs Magn\'{e}tiques
Puls\'{e}s, CNRS/INSA/UPS, UMR 5147, 31400 Toulouse, France.}
\author{M. Nardone}
\affiliation{Laboratoire National des Champs Magn\'{e}tiques
Puls\'{e}s, CNRS/INSA/UPS, UMR 5147, 31400 Toulouse, France.}
\author{R. Battesti}
\affiliation{Laboratoire National des Champs Magn\'{e}tiques
Puls\'{e}s, CNRS/INSA/UPS, UMR 5147, 31400 Toulouse, France.}

\author{L. Martin}
\affiliation{Laboratoire pour l'Utilisation des Lasers Intenses, UMR
7605 CNRS-CEA-X-Paris VI, 91128 Palaiseau, France.}
\author{A.-M. Sautivet}
\affiliation{Laboratoire pour l'Utilisation des Lasers Intenses, UMR
7605 CNRS-CEA-X-Paris VI, 91128 Palaiseau, France.}
\author{J.-L. Paillard}
\affiliation{Laboratoire pour l'Utilisation des Lasers Intenses, UMR
7605 CNRS-CEA-X-Paris VI, 91128 Palaiseau, France.}
\author{F. Amiranoff}
\affiliation{Laboratoire pour l'Utilisation des Lasers Intenses, UMR
7605 CNRS-CEA-X-Paris VI, 91128 Palaiseau, France.}

%
%


\date{\today}

\begin{abstract}
Recently, axion-like particle search has received renewed
interest, and several groups have started experiments. In this paper, we present the final results of our experiment on
photon-axion oscillations in the presence of a magnetic field, which took place at LULI (Laboratoire pour l'Utilisation des Lasers Intenses, Palaiseau, France). Our null measurement allowed us to exclude the existence of axions with inverse coupling constant $M>9.\times 10^5$ GeV for low axion masses and to improve the preceding BFRT limits by a factor 3 or more for axion masses $1.1\, \mbox{meV} <m_a<2.6\,\mbox{meV}$. We also show that our experimental results improve the existing limits on the parameters of a low mass hidden-sector boson usually dubbed ``paraphoton'' because of its similarity with the usual photon. We detail our apparatus which is based on the ``light
shining through the wall'' technique. We compare our results to
other existing ones.
\end{abstract}


\maketitle

\section{Introduction}
\label{sec:Intro}

Ever since the Standard Model was built, various theories have been proposed to go beyond it. Many of these involve, if not imply for the sake of consistency, some light, neutral, spinless particles very weakly coupled to standard model particles, hence difficult to detect.

One famous particle beyond the Standard Model is the axion. Proposed more than 30 years ago to solve the strong CP problem \cite{Peccei1977,Weinberg_Wilczek}, this neutral, spinless, pseudoscalar particle has not been detected yet, in spite of constant experimental efforts \cite{BNL1993,ADMX,CAST2007,Raffelt2007Review}. Whereas the most sensitive experiments aim at detecting axions of solar or cosmic origin, laboratory experiments including the axion source do not depend on models of the incoming axion flux. Because the axion is not coupled to a single photon but to a two-photon vertex, axion-photon conversion requires an external electric or - preferentially - magnetic field to provide for a virtual second photon \cite{Sikivie1983}.

At present, purely terrestrial experiments are built according to two main schemes. The first one, proposed in 1979 by Iacopini and Zavattini \cite{Iacopini1979}, aims at measuring the ellipticity induced on a linearly polarized laser beam by the presence of a transverse magnetic field, but is also sensitive to the ellipticity and, slightly modified, to the dichroism induced by the coupling of low mass, neutral, spinless bosons with laser beam photons and the magnetic field \cite{Maiani1986}. The second popular experimental scheme, named ``light shining through the wall'' \cite{VanBibber1987}, consists of first converting incoming photons into axions in a transverse magnetic field, then blocking the remaining photonic beam with an opaque wall. Behind this wall with which the axions do not interact, a second magnetic field region allows the axions to convert back into photons with the same frequency as the incoming ones. Counting these regenerated photons, one can calculate the axion-photon coupling or put some limits on it. This set-up was first realized by the BFRT collaboration in 1993 \cite{BNL1993}.

Due to their impressive precision, optical experiments relying on couplings between photons and these hidden-sector particles seem most promising. Thanks to such couplings, the initial photons oscillate into the massive particle to be detected. The strength of optical experiments lies in the huge accessible dynamical range: from more than $10^{20}$ incoming photons, one can be sensitive to 1 regenerated photon!

In fact, the ``light shining through the wall'' experiment also yields some valuable information on another hidden-sector hypothetical particle \cite{Popov1991}. After the observation of a deviation from blackbody curve in the cosmic background radiation \cite{Woody1979}, some theoretical works suggested photon oscillations into a low mass hidden sector particle as a possible explanation \cite{Glashow1983}. The supporting model for such a phenomenon is a modified version of electrodynamics proposed in 1982 \cite{Okun1982}, based on the existence of two U(1) gauge bosons. One of the two can be taken as the usual massless photon, while the second one corresponds to an additional massive particle usually called paraphoton. Both gauge bosons are coupled, giving rise to photon-paraphoton oscillations. Several years later, more precise observations did not confirm any anomaly in the cosmic background radiation spectrum \cite{COBERocket1990} and the interest for paraphoton decreased, although its existence was not excluded. More recently, it was found out that similar additional U(1) gauges generally appear in string embeddings of the standard model \cite{StringReviews}, reviving the interest for experimental limits on the paraphoton parameters \cite{Ahlers2007,Jaeckel2008,Ahlers2008}.

Some limits on the mass and the coupling constant of the paraphoton have already been obtained by a photoregeneration experiment \cite{BNL1993}. Astrophysical limits on paraphoton parameters also exist. They have been derived from the agreement of the cosmic microwave background with the blackbody radiation \cite{Bernardis1984}, and more recently by the absence of distortions in the optical spectrum of distant Type Ia supernovae \cite{Angelis2002}.

Our effort was motivated by the observation published by the PVLAS collaboration, and subsequently retracted \cite{Zavattini2007},  which they claimed could be explained by the existence of axions in the mass range 1-2 meV. We have therefore designed an apparatus optimised for that mass region to rapidly prove or disprove the interpretation in terms of axion-like particles of the PVLAS signal. Our preliminary results, excluding at a $3\sigma$ confidence level the existence of axions with parameters consistent with the PVLAS observation, have been published in November 2007 \cite{Robilliard2007}. This paper is devoted to the final results of our ``light shining through the wall'' experiment, sensitive to axion-like particles and to paraphotons. We first derive the detection probabilities for both particles. We then detail our apparatus which strength lies in pulsed laser and magnetic field, thus reducing the demand on the detector noise. Finally, we present our latest experimental results and compare them with the limits obtained by other searches.

\section{Photoregeneration Probability}

\subsection{Axion-Like Particle}

The photon to axion-like particle conversion and reconversion transition probability (in natural units $\hbar=c=1$, with 1 T $\equiv 195$\,eV$^2$ and 1~m $\equiv 5\times 10^6$\,eV$^{-1}$) after propagating over a distance $z$ in the inhomogeneous magnetic field $B$ writes \cite{Sikivie1983,Raffelt1988}:

\begin{equation}
p_\mathrm{a}\left(z\right) = \left| \int_0^z dz' \Delta_M\left(z'\right) \times \exp(i \Delta_\mathrm{a} z')
\right|^2, \label{eq:axion_p_integral}
\end{equation}

\noindent where $\Delta_M = \frac{B}{2M} \quad \mbox{and}\quad \Delta_\mathrm{a} = - \frac{m_\mathrm{a}^2}{2\omega}$, $\omega$ being the photon energy, $m_a$ the axion-like particle mass and $M$ its inverse coupling constant with two photons. Note that this equation is valid for a light polarization parallel to the magnetic field since the axion has to be a pseudoscalar \cite{Peccei1977}. Finally, as we have two identical magnets, the photon regeneration probability due to axion-like particles is

\begin{equation}
P_\mathrm{a} = p_\mathrm{a}^2 \left( L\right),
\label{eq:axion_regenerate}\end{equation}

\noindent with $L$ the magnet length.

In order to have a number of regenerated photons as large as possible, the number of incident photons, the detection
efficiency and the integral of the transverse magnetic field over the magnet length $L$ have to be maximized. We define $B_0$ as the maximum field and $L_{\mathrm{eq}}$ as the equivalent length of a magnet producing a uniform magnetic field $B_0$ such that

\begin{equation}
\int_{-L/2}^{+L/2} B dz = B_0 L_{\mathrm{eq}}.
\end{equation}

\noindent On the other hand, $p_\mathrm{a}(z)$ oscillates for too long magnets. Actually, for a homogeneous magnetic field $B_0$,
Eq. (\ref{eq:axion_p_integral}) gives:

\begin{equation}
p_\mathrm{a} = \left( \frac{B_0L}{2M} \right)^2 \frac{\sin^2(\frac{\Delta_\mathrm{osc}}{2}L)}{(\frac{\Delta_\mathrm{osc}}{2}L)^2},
\label{eq:axion_p}
\end{equation}

\noindent where $\Delta_\mathrm{osc} = -\Delta_\mathrm{a}$. In our case, our search was focused on the $1 \, \mbox{meV} < m_a<2\, \mbox{meV}$, so that a length larger than 1 m would have been useless.

Finally, very recently a detailed theoretical study of the photon to axion-like particle
conversion probability pointed out that an enhancement of this
probability is predicted at $m_\mathrm{a}=\omega$ \cite{Adler2008}. In
this particular condition, the probability of getting a photon after
the wall is :
\begin{equation}
    P_\mathrm{a} = \frac{3\beta^4}{16 q m_\mathrm{a}^4}\log\Big(\frac{2 q
    m_\mathrm{a}^4}{\beta^4}\Big), \label{eq:axion_enhancement}
\end{equation}
with $\beta = B_0/M$ and $q = \Delta/\omega$ the quality factor of
the laser source, $\Delta$ being the laser bandwidth.

\subsection{Paraphoton}

In the modified version of electrodynamics developed in 1982 \cite{Okun1982}, the paraphoton weakly couples with the photon through kinetic mixing. Contrary to axion-like particles, photon-paraphoton oscillations are therefore possible without any
external field and are independent on photon polarization.

Recently, the experimental signatures of paraphoton have been discussed in details in Ref. \cite{Ahlers2008}. The conversion probability of a photon into a paraphoton of mass $\mu$ and {\it vice-versa} after a distance $L$ is given by:

\begin{equation}
p_{\gamma}=4\chi^2\sin^2\left(\frac{\mu^2L}{4\omega}\right)
\label{eq:paraphoton_p}
\end{equation}

\noindent where $\chi$ is the photon-paraphoton coupling constant, which arbitrary value is to be determined experimentally. This equation is valid for a relativistic paraphoton satisfying $\mu \ll \omega$.

Comparing Eqs. (\ref{eq:paraphoton_p}) and (\ref{eq:axion_p}), one notes that from a mathematical point of view the two are equivalent, $\mu$ corresponding to $m_a$, and $\chi$ to ${B_0 \omega}\over{M m_a^2}$. This analogy originates from the fact that both formulas describe the same physical phenomenon, {\it i.e.} quantum oscillations of a two level system. Using this mathematical equivalence between paraphoton parameters and axion-like particle parameters, we were able to derive for the enhancement of the paraphoton conversion probability at $\mu =\omega$ a formula equivalent to Eq. (\ref{eq:axion_enhancement}):

\begin{equation}
P_\gamma = \frac{3\chi^4}{16 q}\log\left(\frac{2 q}{\chi^4}\right).
\label{eq:paraphoton_enhancement}
\end{equation}

In the case of a typical photoregeneration experiment, the incoming photons freely propagate for a distance $L_1$ and might oscillate into paraphotons before being stopped by a wall, after which the paraphotons propagate for a distance $L_2$
and have a chance to oscillate back into photons that are detected with efficiency $\eta_\mathrm{det}$. The photon regeneration probability due to paraphotons can therefore be written as:

\begin{eqnarray}
P_{\gamma} & = & p_{\gamma}(L_1)p_{\gamma}(L_2) \nonumber \\
& = & 16\chi^4 \sin^2\left(\frac{\mu^2L_1}{4\omega}\right) \sin^2\left(\frac{\mu^2L_2}{4\omega}\right)
\label{eq:paraphoton_regenerate}
\end{eqnarray}

\noindent In our experiment, $L_1$ is the distance between the focusing lens at the entrance of the vacuum system, which focuses photons but not paraphotons, and the wall, which blocks photons only. Similarly, $L_2$ represents the distance separating the blind flange just before the regenerating magnet and the lens coupling the renegerated photons into the optical fibre (see Fig. \ref{fig:Setup}).

Note that Eq. (\ref{eq:paraphoton_regenerate}) is {\it a priori} valid in the absence of magnetic field. If a magnetic field is applied, the formula remains valid provided that it can be considered as static during the experiment and its transverse spatial extent is larger than $1/\mu$ \cite{Ahlers2007}, which is the case in our experiment for paraphoton masses larger than $2\times 10^{-5}$\,eV.

\section{Experimental Setup}

\begin{figure}
\begin{center}
\resizebox{1\columnwidth}{!}{
\includegraphics{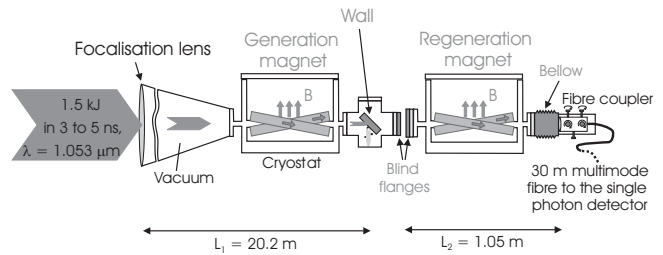}
} \caption{Sketch of the apparatus. The wall and the blind flanges are removable for fibre alignment.} \label{fig:Setup}
\end{center}
\end{figure}

As shown in Fig.\,\ref{fig:Setup}, the experimental setup consists of two
main parts separated by the wall. An intense laser beam travels through a
first magnetic region (generation magnet) where photons might be
converted into axion-like particles. The wall blocks every incident
photon while axion-like particles would cross it without interacting and may be converted back
into photons in a second magnetic region (regeneration magnet).
The regenerated photons are finally detected by a single photon
detector.

The three key elements leading to a high detection rate are the
laser, the generation and regeneration magnets placed on each side
of the wall and the single photon detector. Each element is
described in the following sections.

\subsection{Laser}

In order to have the maximum number of incident photons at a wavelength that can be efficiently detected, the
experiment has been set up at Laboratoire pour l'Utilisation des Lasers Intenses (LULI) in Palaiseau, on the Nano 2000
chain \cite{webLULI}. It can deliver more than 1.5\,kJ over a few nanoseconds with $\omega =
1.17$\,eV. This corresponds to $N_i = 8\times10^{21}$\,photons per
pulse.

The nanosecond pulse is generated by a YLF seeded oscillator with a
$\Delta = 1.7$\,meV bandwidth. It delivers 4\,mJ with a duration
adjustable between 500\,ps and 5\,ns. Temporal shaping is realized
with five Pockels cells. Then this pulse seeds single-pass
Nd:Phosphate glass rods and disk amplifiers. During our 4 weeks of
campaign, the total duration was decreased from 5\,ns the first week
to 4\,ns and finally 3\,ns while keeping the total energy constant.
A typical time profile is shown in the inset of
Fig.\,\ref{fig:B_time} with a full width at half maximum of 2.5\,ns
and a total duration of 4\,ns.

\begin{figure}
\begin{center}
\resizebox{1\columnwidth}{!}{
\includegraphics{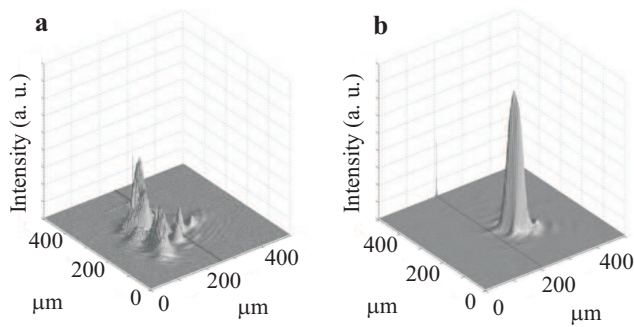}
} \caption{Focal spot without correction(a) and with wave-front
correction (b). This correction allows to maintain a spot of one or
two diffraction limits despite the amplifiers' not being in thermal
equilibrium.} \label{fig:Deformable_mirror}
\end{center}
\end{figure}

The repetition rate of high energy pulses is imposed by the
relaxation time of the thermal load in the amplifiers which implies
wave-front distortions. Dynamic wave-front correction is applied by
use of an adaptive-optics system \cite{DeformableMirror}. To this
end a deformable mirror is included in the middle of the
amplification chain. It corrects the spatial phase of the beam to
obtain at focus a spot of about once or twice the diffraction limit, as
shown in Fig.\,\ref{fig:Deformable_mirror}. This system allows to
increase the repetition rate while maintaining good focusability
although the amplifiers are not at thermal equilibrium. During data
acquisition, the repetition rate has typically varied between 1
pulse per hour and 1 pulse every other hour.

At the end of the amplification chain, the vertically linearly
polarized incident beam has a 186\,mm diameter and is almost
perfectly collimated. It is then focused using a lens which focal
length is 20.4\,m. The wall is placed at $L_1 = 20.2$\,m from the
lens in order to have the focusing point a few centimeters behind this wall. The
beam is well apodized to prevent the incoming light from generating
a disturbing plasma on the sides of the vacuum tubes.

Before the wall where the laser beam propagates, a vacuum better
than $10^{-3}$\,mbar is necessary in order to avoid air ionization.
Two turbo pumps along the vacuum line easily give $10^{-3}$\,mbar
near the lens and better than $10^{-4}$\,mbar close to the wall. The
wall is made of a 15\,mm width aluminum plate to stop every incident
photon. It is tilted by 45\,$^\circ$ with respect to the laser beam axis in order to increase the area of the laser impact
and to avoid retroreflected photons. In the second magnetic field
region, a vacuum better than $10^{-3}$\,mbar is also maintained.

\begin{figure}
\begin{center}
\resizebox{1\columnwidth}{!}{
\includegraphics{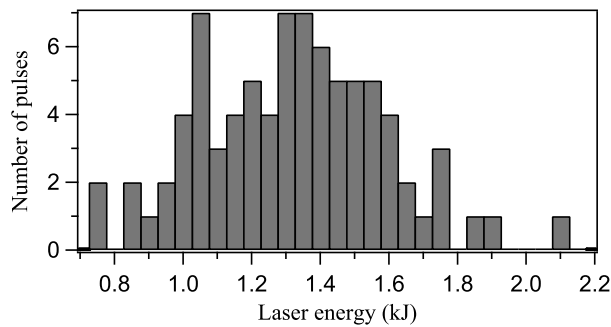}
} \caption{Number of high energy pulses versus laser energy during
the four weeks of data acquisition.} \label{fig:HistEnergy}
\end{center}
\end{figure}

Fig.\,\ref{fig:HistEnergy} shows a histogram of laser energy per
pulse for the 82 laser pulses performed during our campaign. The laser energy per pulse ranges from 700\,J to 2.1\,kJ, with a mean value of 1.3\,kJ.

\subsection{Magnetic field}

\begin{figure}
\begin{center}
\resizebox{1\columnwidth}{!}{
\includegraphics{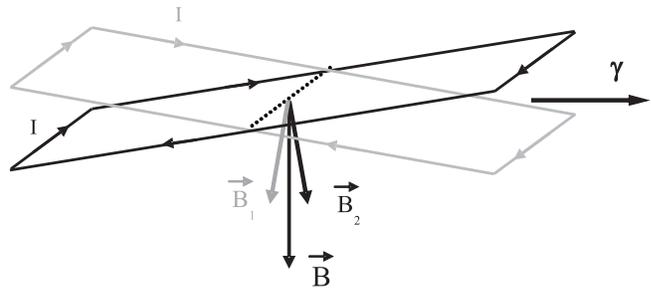}
} \caption{Scheme of XCoil. Magnetic fields $\vec{B_1}$ and
$\vec{B_2}$ are created by each of the race-track shaped windings. This
yields a high transverse magnetic field $\vec{B}$ while allowing
the necessary optical access for the laser photons $\gamma$.}
\label{fig:XCoil_Scheme}
\end{center}
\end{figure}

Concerning the magnets, we use a pulsed technology. The pulsed magnetic field
is produced by a transportable generator developed at LNCMP \cite{Frings2006}, which
consists of a capacitor bank releasing its energy in the coils in a
few milliseconds. Besides, a special coil geometry has been
developed in order to reach the highest and longest transverse
magnetic field. Coil properties are explained in
Ref.\,\cite{Batut2008}. Briefly, the basic idea is to get the
wires generating the magnetic field as close as possible to the
light path. As shown in Fig.\,\ref{fig:XCoil_Scheme}, the coil consists of
two interlaced race-track shaped windings that are tilted one with
respect to the other. This makes room for the necessary optical
access at both ends in order to let the laser in while providing a
maximum $B_0 L_{\mathrm{eq}}$. Because of the particular arrangement
of wires, these magnets are called Xcoils.

The coil frame is made of G10 which is a non conducting material
commonly used in high stress and cryogenic temperature conditions.
External reinforcements with the same material have been added after
wiring to contain the magnetic pressure that can be as high as
500\,MPa. A 12\,mm diameter aperture has been dug into the magnets
for the light path.

As for usual pulsed magnets, the coils are immersed in a liquid
nitrogen cryostat to limit the consequences of heating. The whole
cryostat is double-walled for a vacuum thermal insulation. This
vacuum is in common with the vacuum line and is better than
$10^{-4}$\,mbar. A delay between two pulses is
necessary for the magnet to cool down to the equilibrium temperature which is
monitored via the Xcoils' resistance. Therefore, the repetition
rate is set to 5 pulses per hour. Furthermore the coils' resistance is
precisely measured after each pulse and when equilibrium is reached,
in order to check the Xcoils' non embrittlement. Indeed variations
of the resistance provide a measurement of the accumulation of defects
in the conductor material that occur as a consequence of plastic
deformation. These defects lead to hardening and embrittlement of
the conductor material, which ultimately leads to failure.

\begin{figure}
\begin{center}
\resizebox{1\columnwidth}{!}{
\includegraphics{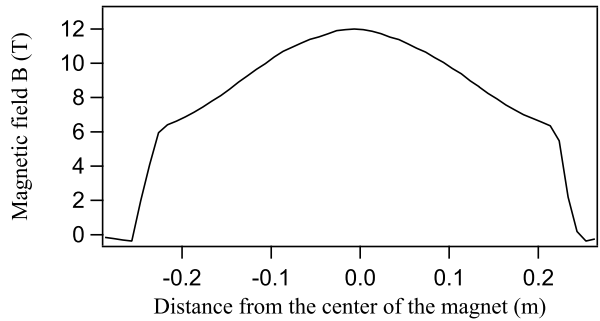}
} \caption{Transverse magnetic field inside the magnet along the
laser direction. At the center of the magnet we have a mean maximum
magnetic field $B_0 = 12$\,T. Integrating $B$ along the optical path
yields 4.38\,T.m.} \label{fig:B}
\end{center}
\end{figure}

The magnetic field is measured by a calibrated pick-up coil. This
yields the spatial profile shown in Fig.\,\ref{fig:B}. The maximum
field $B_0$ is obtained at the center of the magnet. Xcoils have
provided $B_0 \geq 13.5$\,T over an equivalent length
$L_{\mathrm{eq}} = 365$\,mm. However, during the whole
campaign a lower magnetic field of $B_0=12\, (0.3)$\,T was
used to increase the coils' lifetime.

\begin{figure}
\begin{center}
\resizebox{1\columnwidth}{!}{\includegraphics{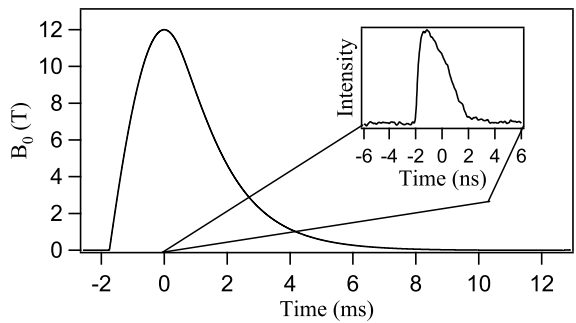}}
\caption{Magnetic field $B_0$ at the center of the magnet as a
function of time. The maximum is reached within 1.75\,ms and can be
considered as constant ($\pm 0.3\%$) during $\tau_{B} = 150\,\mu$s.
The 3 to 5\,ns laser pulse is applied during this interval. Inset:
temporal profile of a 4\,ns laser pulse.} \label{fig:B_time}
\end{center}
\end{figure}

A typical time dependence of the pulsed magnetic field at the center
of the magnet is represented in Fig.\,\ref{fig:B_time}. The total
duration is a few milliseconds. The magnetic field reaches its maximum value within less than
2\,ms and remains constant ($\pm 0.3\%$) during $\tau_{B} =
150\,\mu$s, a very long time compared to the laser pulse.

\subsection{Detector}

The last key element is the detector that has to meet several
criteria. In order to have a sensitivity as good as possible, the
regenerated photon detection has to be at the single photon level.
The integration time is limited by the longest duration of the laser
pulse which is 5\,ns. Since we expected about 100 laser pulses during our four week campaign,
which corresponds to a total integration time of $500$\,ns, we required a detector with a dark count rate \footnote{A dark count, originating from electronic noise,
corresponds to the apparent detection of a photon while no light strikes the detector.} far lower than 1
over this integration time, so that any increment of the counting
would be unambiguously associated to the detection of one
regenerated photon.

Our detector is a commercially available single photon receiver from
Princeton Lightwave which has a high detection efficiency at
$1.05\,\mu$m. It integrates a $80\times80\,\mu$m$^2$ InGaAs
Avalanche Photodiode (APD) with all the necessary bias, control and
counting electronics. Light is coupled to the photodiode through a
FC/PC connector and a multimode fiber. When the detector is
triggered, the APD bias voltage is raised above its reverse
breakdown voltage $V_{\mathrm{br}}$ to operate in ``Geiger mode''. A
short time later -- adjustable between 1\,ns and 5\,ns -- the bias
is reduced below $V_{\mathrm{br}}$ to avoid false events. For our
experiment, the bias pulse width is 5\,ns to correspond with the
longest laser pulse.

\begin{figure}
\begin{center}
\resizebox{1\columnwidth}{!}{\includegraphics{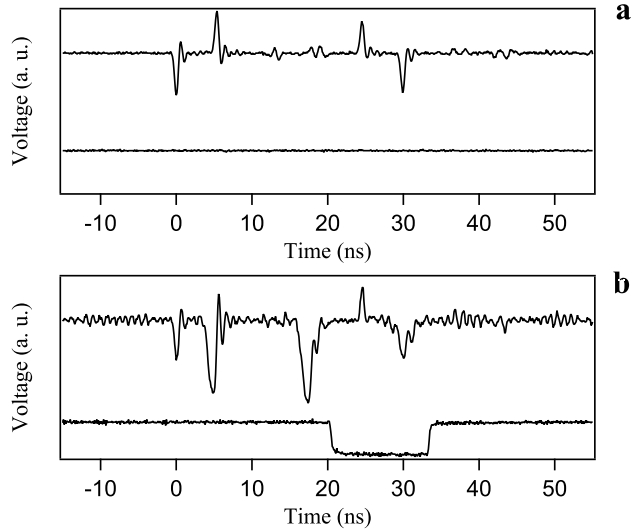}}
\caption{Amplified APD output (upper curve) and logic signal (lower
curve) of the detector as a function of time. The capacitive
transients on the APD output signals are due to the gated
polarisation of the photodiode in Geiger mode. (a) Signals with no
incident photon. (b) Signals when a photon is detected.}
\label{fig:SignalDet}
\end{center}
\end{figure}

Typical output signals available on the detector are plotted in
Fig.\,\ref{fig:SignalDet}. Let's first consider
Fig.\,\ref{fig:SignalDet}a with no incident photon. The upper signal
corresponds to the amplified APD output. The application of such
a short pulse to a reverse-biased APD produces a capacitive transient.
The first two transients temporally shifted by 5\,ns correspond to
the bias pulse. This signal enables to precisely determine the
moment when detection starts. The last transients are due to an
electronic reflection of the bias pulse.

When a photon is detected (Fig.\,\ref{fig:SignalDet}b), the signal
resulting from a photon-induced avalanche superimposes upon
transients. The transient component may be much larger than the
photon-induced component, making it difficult to discern. The
detector uses a patented transient cancelation scheme to overcome
this problem \cite{SinglePhotonDetector}. A replica of the unwanted
transient is created and subtracted from the initial signal. The
photon-induced signal will thus appear against a flat, low-noise
background, as it is observed in Fig.\,\ref{fig:SignalDet}b between
the initial bias pulse and the reflected one. It can then be easily
detected using a discriminator. To this end, this signal is sent to
a fast comparator with adjustable threshold that serves as a
discriminator and outputs a logic pulse, as shown by lower traces on
Fig.\,\ref{fig:SignalDet}.

\begin{figure}
\begin{center}
\resizebox{1\columnwidth}{!}{\includegraphics{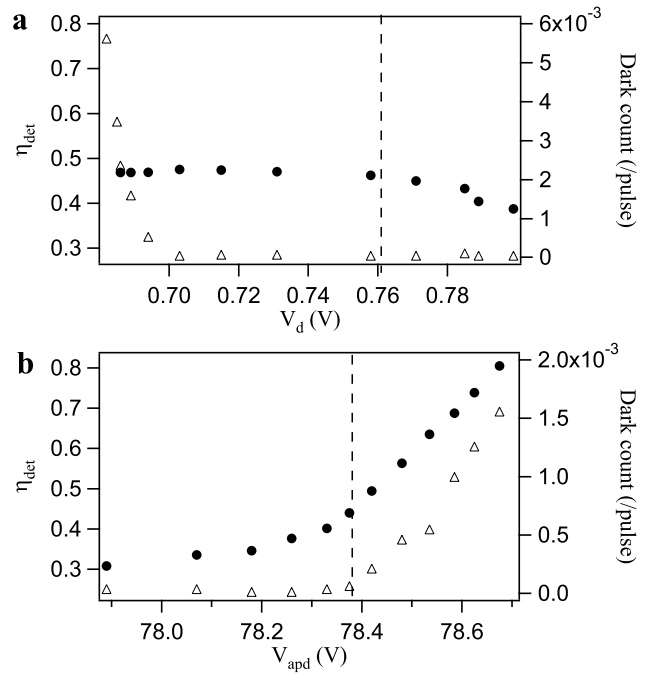}}
\caption{Detection efficiency ($\bullet$) and dark count per 5\,ns
bias pulse ({\tiny $\triangle$}) as a function of the discriminator
threshold (a) ($V_{\mathrm{APD}}$ fixed to 78.4\,V) or APD bias
voltage (b) ($V_{\mathrm{d}}$ fixed to 0.760\,V). The APD
temperature is fixed to the lowest achievable value 221.5\,K. Dashed
lines indicate the chosen working point.} \label{fig:Vapd}
\end{center}
\end{figure}

To optimize the dark count rate and the detection efficiency
$\eta_\mathrm{det}$, three different parameters can be adjusted:
the APD temperature, the discriminator threshold $V_\mathrm{d}$ set to
reject electronic noise and the APD bias voltage $V_{\mathrm{APD}}$.
The dark count rate is first optimized by choosing the lowest
achievable temperature which is around 221\,K. This rate is
measured with no incident light, a trigger frequency of 5\,kHz and
an integration time of at least 1\,s. Dark counts for a 5\,ns
detection gate as a function of $V_\mathrm{d}$ is shown in
Fig.\,\ref{fig:Vapd}a. It increases rapidly when
$V_\mathrm{d}$ is too low. On the other hand, $\eta_\mathrm{det}$
remains constant for a large range of $V_\mathrm{d}$. We set
$V_\mathrm{d}$ to a value far from the region where dark count
increases and where $\eta_\mathrm{det}$ is still constant. This
corresponds to less than $2.5\times10^{-2}$ dark count over
$500$\,ns integration time.

\begin{figure}
\begin{center}
\resizebox{1\columnwidth}{!}{\includegraphics{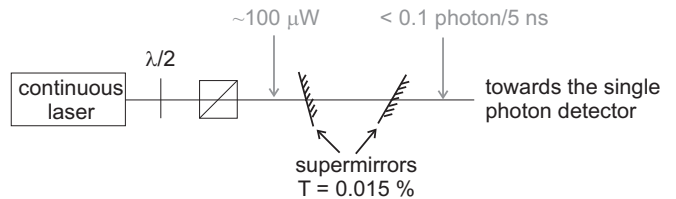}}
\caption{Experimental setup to measure the detection efficiency of
the single photon detector. The detector is illuminated with a laser
intensity lower than 0.1 photon per 5\,ns. This intensity is
calculated through the measurement of the supermirrors transmission
and the laser power before those supermirrors. An half waveplate and
a polarizer are used to change the number of incident photons.}
\label{fig:SinglePhoton}
\end{center}
\end{figure}

The detection efficiency is precisely measured by illuminating the
detector with a laser intensity lower than 0.1 photon per detection
gate at 1.05\,$\mu$m. The probability to have more than one photon
per gate is thus negligible. Such a low intensity is obtained with
the setup described in Fig. \ref{fig:SinglePhoton}. A c.w.
laser is transmitted through two supermirrors with a reflectivity
higher than 99.98\,$\%$ \footnote{The main advantage of using mirrors
to strongly decrease the laser intensity instead of densities is to avoid thermal effects within the optics and thus to obtain a
transmission independent on incident power.}. The angle of incidence
is near normal in order to intercept the reflected beam and avoid
spurious light without increasing transmission. This gives a
measured transmission of $0.015\,\%$ for each mirror. Finally, to
calculate the number of incident photons on the detector, we measure
the laser power before the two supermirrors with a precise power
sensor.

The detection efficiency as a function of the bias voltage is
plotted in Fig. \ref{fig:Vapd}b. Our measurements show
that $\eta_\mathrm{det}$ slowly increases with $V_{\mathrm{APD}}$
until a threshold where it increases dramatically for a value of
$V_{\mathrm{APD}}$ shortly below the dark count runaway value. The
best compromise between detection efficiency and dark count rate is
found at $V_{\mathrm{APD}} = 78.4 (0.05)$\,V with $\eta_\mathrm{det}
= 0.48 (0.025)$.\\

As said in the introduction, other similar experiments generally
require long integration times which implies an experimental
limitation due to the detection noise. Using pulsed laser,
magnetic field and detection is an original and efficient
way to overcome this problem. Photons are concentrated in very
intense short laser pulses during which the detection background is
negligible. This also means that if a photon is detected in our
experiment in correlation with the magnetic field, it will be an unambiguous signature of axion generation
inside our apparatus.

\section{Experimental protocol and tests}

\subsection{Alignment}

After the second magnet, the regenerated photons are injected into the
detector through a coupling lens and a graded index multimode fiber
with a 62.5\,$\mu$m core diameter, a 0.27 numerical aperture and an
attenuation lower than 1\,dB/km. These parameters ensure that we can
inject light into the fiber with a high coupling ratio, even
when one takes into account the pulse by pulse instability of the
propagation axis that can be up to 9\,$\mu$rad.

Injection is adjusted thanks to the fiber coupler, and by removing
the wall and the blind flanges (see Fig. 1). As the high energy
laser beam, the alignment beam comes from the pilot oscillator
without chopping nor amplifying it. This procedure ensures that the
pulsed kJ beam is perfectly superimposed to the alignment beam.
During data acquisition, the mean coupling efficiency through the
fibre was found to be $\eta_c = 0.85$.

The alignment of the high energy beam is performed with a low energy 5\,ns pulsed beam, allowing for a 10\,Hz repetition rate. During alignment, several black crosses
are distributed along the laser path to mark the optical axis.
Mirrors mounted on stepper motors allow to align the
beam very precisely on this axis. This procedure is carried out a few minutes before each high
energy pulse.

\begin{figure}
\includegraphics[width=8cm]{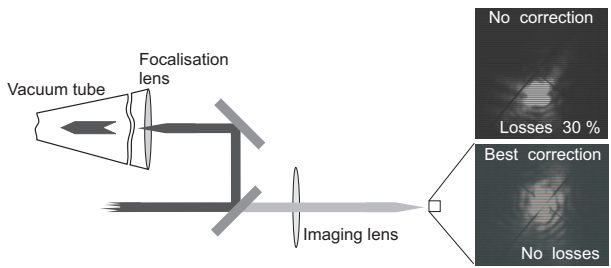}\\
\caption{Monitoring of the optical path followed by the high energy
beam. Losses due to misalignment are estimated by comparing the
centre of the beam to the centre of the black cross. The upper image corresponding to an uncorrected laser beam pointing exhibits 30\% injection losses, while the lower one is perfectly corrected.}\label{fig:EnergyPath}
\end{figure}

The only remaining source of misalignment lies in thermal effects
during the high energy pulse, which could slightly deviate the laser
beam, hence generating supplementary losses in fibre coupling. This
misalignment is mostly reproducible. This means that it can be corrected by
a proper offset on the initial laser pointing. The far field of the high energy beam is imaged for each pulse at the output of the amplification chain (see Fig. \ref{fig:EnergyPath}). Since the focal length of the imaging system is similar to that of our focalisation lens, the position of the far field image on the alignment mark is a fair diagnosis of the alignment on the fiber coupler. The best offset was determined by trial and error method after a few high energy pulses.

\subsection{Optical and electro-magnetic noise}

In order to have the best sensitivity, a perfect optical shielding
is necessary. As shown in Fig.\,\ref{fig:Setup}, an aluminum blind flange
closes the entrance to the regeneration magnet. A black soft PVC bellow placed between the exit of the magnet and the fibre coupler prevents stray light while mechanically decoupling the
magnet which vibrates during its pulse and the fibre coupler which
should stay perfectly still. Finally, another aluminum blind flange
closes the exit of the generation area in order to stop any incident
photon scattered inside the vacuum line.

A count on the single-photon receiver is most probably due to an
incident photon on the photodiode but it may also originate from
electro-magnetic noise during laser or magnetic pulses. To avoid
such noise, the detector is placed in a Faraday shielding bay. In addition,
a 30\,m long fibre is used so that the detector can be placed far away
from the magnets.

To test our protective device, laser and magnetic pulses were
separately applied while triggering the detector. No fake signal was
detected, validating the optical and electro-magnetic shielding.

\subsection{Synchronization}

Our experiment is based on pulsed elements which require a perfect
synchronization : the laser pulse must cross the magnets when the
magnetic field is maximum and fall on the photodiode during the detection gate.

The magnetic pulse is triggered with a TTL signal from the laser chain.
The delay between this signal and the laser trigger is adjusted once
and for all by monitoring on the same oscilloscope the magnetic
field and the laser trigger. Then, the magnetic trigger has a
jitter lower than 10\,$\mu$s, ensuring that the laser pulse travels through the magnets within the
150\,$\mu$s interval during which the magnetic field is constant and maximum.

\begin{figure}
  \includegraphics[width=8cm]{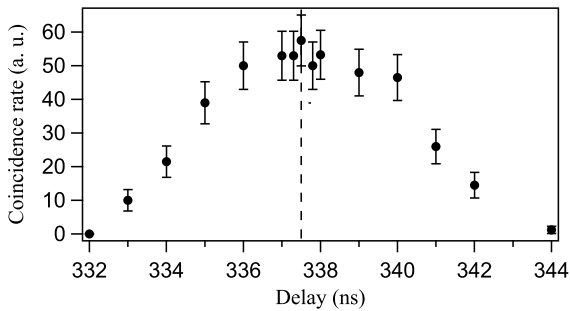}\\
  \caption{Coincidence rate between the arrival of photons on the
detector and its 5\,ns detection gate as a function of an arbitrary
delay time. The dashed line indicates our working point, chosen in
order to maximize the coincidence rate.}\label{fig:Coinc}
\end{figure}

Synchronization of the laser pulse and the detector needs to be far more
accurate since both have a 5\,ns duration. The detector gate is triggered with the same fast signal
as the laser, using delay lines. We have measured the coincidence
rate between the arrival of photons on the detector and the opening
of the 5\,ns detector gate as a function of an adjustable delay. We
have chosen our working point in order to maximize the coincidence
rate (see Fig. \ref{fig:Coinc}). To perform such a measurement we used
the laser pilot beam which was maximally attenuated by shutting off 4 Pockels cells along the amplification chain and chopped with
a pulsed duration of 5\,ns, which corresponds to the longest
duration of the kJ beam.

\section{Data analysis}

\subsection{Detection sensitivity}

The best experimental limits are achieved when no fake signal is detected during the experiment. In this case, to estimate
the corresponding upper conversion probability of regenerated
photons, we have to calculate the upper number of photons that could
have been missed by the detector for a given confidence level ($CL$).

The probability $P_n$ that $n$ incident photons have been missed by
the detector is $P_n = (1-\eta_\mathrm{det})^n$ when dark count is
negligible. Therefore, the probability that $n$ photons at most were missed by the detector writes

$$\frac{\sum_{k=0}^{n}P_k} {\sum_{k=0}^{\infty}P_k} =1-\left( 1-\eta_\mathrm{det}\right)^{n+1}$$

\noindent and has to be compared with the required confidence level $CL$. This yields the upper number of possibly missed photons $n_\mathrm{missed}$ as the smallest integer $n$ satisfying
$$1-\left(1-\eta_\mathrm{det}\right) ^{n+1} \geq CL,$$

\noindent which writes

\begin{equation}
n_\mathrm{missed} = \frac{\log(1-CL)}{\log(1-\eta_\mathrm{det})}-1.
\label{eq:n_missed}
\end{equation}

\noindent For example, with our value of $\eta_\mathrm{det}$, a confidence
level of 99.7\,$\%$ corresponds to less than 8 missed photons. The
upper photon regeneration probability is then
\begin{equation}
P_{\mathrm{a}\;\mathrm{or}\;\gamma} = \frac{n_\mathrm{missed}}{N_{\mathrm{eff}}},
\label{eq:Proba}
\end{equation}
where $N_{\mathrm{eff}}$ is the number of effective incident photons over the
total number of laser shots, taking into account the losses described hereafter. Our experimental sensitivity limit for
the coupling constant versus mass is finally calculated by
numerically solving Eqs. (\ref{eq:axion_p_integral}) and
(\ref{eq:axion_regenerate}) for axion-like particles, and
Eqs. (\ref{eq:paraphoton_p}) and (\ref{eq:paraphoton_regenerate})
for paraphotons.

\subsection{Photon losses}

The number of photons per laser pulse $N_i$ is measured at the end of the amplification chain with a calibrated
calorimeter. Then the number of effective incident photons on the detector
$N_{\mathrm{eff}}$ should take into account every losses.  The first
source of losses is due to the coupling efficiency through the
fibre. This is precisely calibrated once a day. Injection is checked before each pulse, just after the alignement of the high energy beam.
The mean coupling efficiency is $\eta_\mathrm{c} = 0.85$.

As said before, the main source of misalignment lies in
thermal effects during the high energy laser pulse, which mean value was corrected. Furthermore, using the c.w. alignment beam we calibrated the injection losses in the fibre as a function of the misalignment visible on the far field imaging. Thanks to this procedure, we were able to estimate the actual alignment losses for each pulse: they amounted to 30\,\% for a non-corrected pulse and varied between 0 and 10\,\% for corrected pulses, because of pulse-to-pulse instabilities.

Possible jitter between the beginning of the detection and the
arrival of the laser pulse on the detector is also taken into
account. For each pulse, a single oscilloscope acquires the laser
trigger, the detector trigger as well as the detection gate. Those
curves allow to precisely calculate the moment $t_0$ when detection
actually starts compared to the laser pulse arrival. Furthermore, the
temporal profile of each laser pulse is also monitored. By
integrating this signal from $t_0$ and during the 5\,ns of detection,
the fraction $\eta_\mathrm{f}$ of photons inside the detection gate
is calculated. This fraction has fluctuated between 0.6 and 1 at the
beginning of our data acquisition with the 5\,ns pulse, mainly due a
1\,ns jitter that was then reduced to about 200\,ps. Then, with the 4\,ns and 3\,ns laser pulses, jitter is
less critical and $\eta_\mathrm{f} = 1$ is obtained almost all the
time.

Finally, for axion-like particles the numerical solving of
Eq.(\ref{eq:axion_p_integral}) is performed with a fixed magnetic field
$B_0$. Variations of this magnetic field along data acquisition are
taken into account by multiplying each number of incident photons by
the factor $(B_{0,i}/B_0)^4$, where $B_{0,i}$ is the maximum field
for the $i^\mathrm{th}$ pulse.

Integration of every losses yields a total number of effective photons
\begin{equation}
N_\mathrm{eff,\, a} = \sum_i \eta_{\mathrm{c},i} \eta_{\mathrm{p},i}
\eta_{\mathrm{f},i} \Big(\frac{B_{0,i}}{B_0}\Big)^4 N_i,
\label{eq:axion_Neff}
\end{equation}

\noindent the sum being taken over the total number of laser and magnetic pulses.

Concerning paraphotons, given that the magnetic fields has no effect on the oscillations, the formula writes

\begin{equation}
N_\mathrm{eff,\, \gamma} = \sum_i \eta_{\mathrm{c},i} \eta_{\mathrm{p},i} \eta_{\mathrm{f},i} N_i.
\label{eq:paraphoton_Neff}
\end{equation}

\section{Results}

Data acquisition was spread over 4 different weeks. As shown in
Fig.\,\ref{fig:HistEnergy}, 82 high energy pulses have reached the
wall with a total energy of about 110\,kJ. This corresponds to
$5.9\times10^{23}$ photons. During the whole data acquisition, no signal has been detected.

\subsection{Axion-Like Particles}

The magnetic field was applied during 56 of those laser pulses, with
a mean value of 12\,T. The laser pulses without magnetic field aimed at testing for possible fake counts.

\begin{figure}
\begin{center}
\resizebox{1\columnwidth}{!}{
\includegraphics{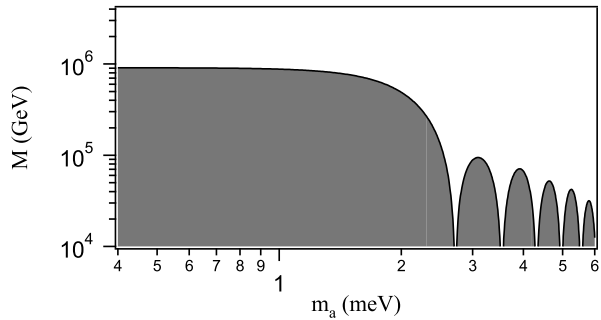}
} \caption{3$\sigma$ limits for the axion-like particle - two photon
inverse coupling constant $M$, as a function of the axion-like
particle mass $m_\mathrm{a}$, obtained from our null result. The area
below our curve is excluded.} \label{fig:Axion_LULI}
\end{center}
\end{figure}

Our experimental sensitivity limits for axion-like particle at
$99.7\,\%$ confidence level are plotted on
Fig.\,\ref{fig:Axion_LULI}. They correspond to a detection
probability of regenerated photons $P_\mathrm{a} =
3.3\times10^{-23}$ and give $M>9.1\times10^5$\,GeV at
low masses. The dark gray area below our curve is excluded. This improves the limits we have published in \cite{Robilliard2007}, which already excluded the PVLAS results \cite{Zavattini2007}.

\begin{figure}
\begin{center}
\resizebox{1\columnwidth}{!}{
\includegraphics{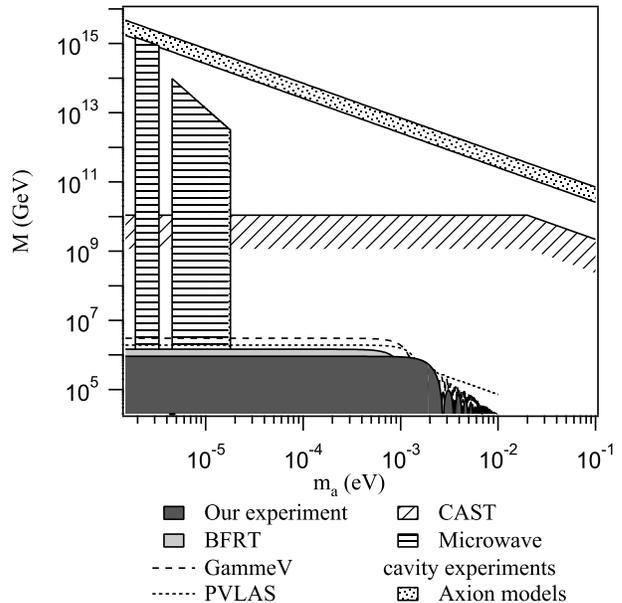}
} \caption{Limits on the axion-like particle - two photon inverse
coupling constant $M$ as a function of the axion-like particle mass
$m_\mathrm{a}$ obtained by experimental searches. Our exclusion
region is first compared to other purely laboratory experiments such as the BFRT photon regeneration experiment
\cite{BNL1993}, the GammeV experiment \cite{FermiLab2008} and the
PVLAS collaboration \cite{PVLAS2008} with a 3$\sigma$ confidence level. Those curves are finally
compared to the 95\,$\%$ confidence level exclusion region obtained
on CAST \cite{CAST2007} and the more than 90\,$\%$ confidence level
on microwave cavity experiments \cite{CavityRBF,CavityUF,ADMX}.
Model predictions are also shown as a dotted stripe between the predictions of the KSVZ model (lower line, $E/N=0$) \cite{KSVZ} and of the DFSZ model (upper line, $E/N=8/3$) \cite{DFSZ}.} \label{fig:CourbeGenerale_Axion}
\end{center}
\end{figure}

We also compared our limits to other laboratory experiments in
Fig.\,\ref{fig:CourbeGenerale_Axion}. They are comparable to other
purely laboratory experiments \cite{BNL1993,FermiLab2008,PVLAS2008},
especially in the meV region of mass. On the other hand, they are
still far from experiments which limits (stripes) approach models
predictions \cite{CavityRBF,CavityUF,ADMX,CAST2007}.

Using Eq. (\ref{eq:axion_enhancement}), our experimental results correspond to $M >
8$\,GeV at $m_\mathrm{a} = 1.17$\,eV. Despite this enhancement, our limits are still very far from the inverse coupling constant of
model predictions which is around $10^9$\,GeV for a 1\,eV mass.

\subsection{Paraphotons}

In the case of paraphotons, we take into account the laser bandwidth $\Delta$ by averaging $P_\gamma (\omega)$ over $\Delta$:

\begin{equation}
\overline{P_\gamma} =
\frac{1}{\Delta}\int_{\omega-\frac{\Delta}{2}}^{\omega-\frac{\Delta}{2}}
P_\gamma(\omega) d\omega.
\end{equation}

\noindent The experimental sensitivity is then calculated by numerically solving

\begin{equation}
\overline{P_\gamma}=\frac{n_\mathrm{missed}}{N_\mathrm{eff}},
\end{equation}

\noindent where $N_\mathrm{eff}$ is given by Eq. (\ref{eq:paraphoton_Neff}). In the regime of low mass $\mu \ll\sqrt{\omega/L q}$, it is equivalent to $\overline{P_\gamma} = P_\gamma$ and the mixing parameter oscillates as a function of the paraphoton
mass. For higher masses, oscillations are smoothed to a mean value. Note that the relevant mass ranges concerning
axion-like particles are situated in the low mass regime, which explains why the averaging over the laser bandwidth was not useful.

\begin{figure}
\begin{center}
\resizebox{1\columnwidth}{!}{
\includegraphics{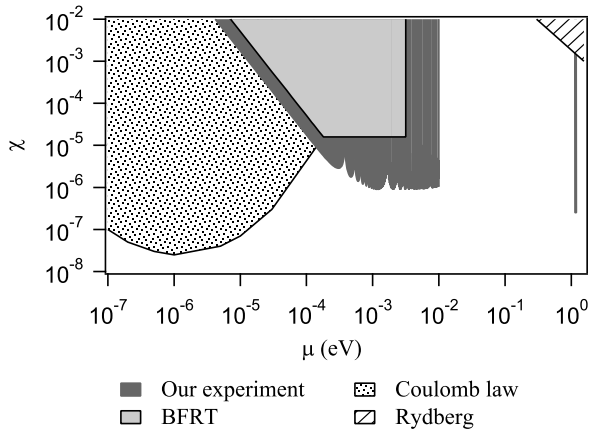}
} \caption{$95\,\%$ confidence level limits on photon-paraphoton
mixing parameter as a function of the paraphoton mass obtained to
our null result (deep gray area). Shaded regions are excluded. This
is compared to excluded regions obtained on BFRT photon regeneration
experiment \cite{BNL1993} (light gray area), to searches for
deviations of the Coulomb law \cite{CoulombLaw} (points) and to
comparisons of the Rydberg constant for different atomic transitions
\cite{Rydberg} (stripes).} \label{fig:CourbeGenerale_Paraphoton}
\end{center}
\end{figure}

The deep gray area in Fig.\,\ref{fig:CourbeGenerale_Paraphoton} represents
the parameters for paraphoton that our measurements exclude with a $95\,\%$ confidence level. It
corresponds to a maximum photon regeneration probability $P_\gamma =
9.4\times10^{-24}$. This sets
a limit $\chi < 1.1\times10^{-6}$ for 1\,meV$<\mu<$10\,meV (for
higher masses, Eq.(\ref{eq:paraphoton_p}) is not valid anymore). This
improves by almost one order of magnitude the exclusion area
obtained on BFRT photon regeneration experiment \cite{BNL1993}. The
enhanced probability at $\mu = \omega$ given by Eq. (\ref{eq:paraphoton_enhancement}) corresponds to $\chi <
1.9\times 10^{-7}$. For other ranges of mass, a more
complicated calculation is required \cite{Adler2008} which is beyond the
scope of this article. Nevertheless, comparing to other laboratory
experiments \cite{CoulombLaw,Rydberg} (see \cite{Review_ParaExp} for
review), we were able to constrain the paraphoton parameters in a region which had not been
covered so far by purely terrestrial experiments.

\section{Conclusion and Outlooks}

We have presented the final results of our photon regeneration experiment which exclude the PVLAS results. Our null measurement leads to limits similar to other purely terrestrial axion searches, and improves the preceding limits by more than one order of magnitude concerning paraphotons \cite{Ahlers2007}.

As far as axion-like particles are concerned, improving the sensitivity of our apparatus in order to test the axion model predictions seems rather unrealistic, especially as the possible mass and two photon coupling constant ranges are still several orders of magnitude wide. In that respect, magnetic birefringence experiments like the one presently under development in Toulouse \cite{BMVprototype} seem more promising: aimed at measuring for the first time the long predicted QED magnetic birefringence of vacuum \cite{QEDBMV}, it will improve by one to two orders of magnitude the precision of purely terrestrial axion searches.

More generally, let us argue that such precision optical experiments may prove useful for experimentally testing the numerous theories beyond standard model in the low energy window, a range in which the large particle accelerators are totally helpless. For example, our apparatus can be modified to become sensitive to chameleon fields \cite{Chameleon}.

Finally, very intense laser beams such as those planned at ELI \cite{ELIwebsite} will become available in the forthcoming years. Such facilities should open new exciting opportunities for our field.

\section*{ACKNOWLEDGMENTS}

We thank the technical staff from LCAR, LNCMP and LULI, especially
S. Batut, E. Baynard,  J.-M. Boudenne, J.-L. Bruneau, D. Castex,
J.-F. Devaud, P. Frings, M. Gianesin, P. Gu\'ehennec, B. Hirardin,
J.-P. Laurent, L. Polizzi, W. Volondat, and A. Zitouni.
We also thank B. Girard and G. Rikken for strongly
supporting this project. This work has been possible thanks to the
ANR-Programme non th\'{e}matique (Contract ANR - BLAN06-3-139634).

\bibliography{apssamp}

\begin{thebibliography}{}


\bibitem{Peccei1977} R. D. Peccei and H. R. Quinn, Phys. Rev. Lett. {\bf 38}, 1440 (1977); R. D. Peccei and H. R. Quinn, Phys. Rev. D {\bf 16}, 1791 (1977).

\bibitem{Weinberg_Wilczek} S. Weinberg, Phys. Rev. Lett. {\bf 40}, 223 (1978); F. Wilczek, Phys. Rev. Lett. {\bf 40}, 279 (1978).

\bibitem{BNL1993} R. Cameron {\it et al.}, Phys. Rev. D {\bf 47}, 3707 (1993).

\bibitem{ADMX} S. J. Asztalos, {\it et al.}, Phys. Rev. D
{\bf{69}}, 011101(R) (2004); L.D. Duffy {\it et al.}, Phys. Rev. D {\bf 74}, 012006 (2006).

\bibitem{CAST2007} S. Andriamonje {\it et al.} (CAST collaboration), J. Cosmol. Astropart. Phys. {\bf 04}, 010 (2007).

\bibitem{Raffelt2007Review} For a review, see G. G. Raffelt, J. Phys. A {\bf 40}, 6607 (2007) and references therein.

\bibitem{Sikivie1983} P. Sikivie, Phys. Rev. Lett. {\bf 51}, 1415 (1983);
P. Sikivie, Phys. Rev. D {\bf 32}, 2988 (1985),

\bibitem{Iacopini1979} E. Iacopini and E. Zavattini, Phys. Lett. B {\bf 85}, 151 (1979).

\bibitem{Maiani1986} L. Maiani, R. Petronzio and E. Zavattini, Phys. Lett B {\bf 175}, 359 (1986).

\bibitem{VanBibber1987} K. Van Bibber {\it et al.}, Phys. Rev. Lett. {\bf{59}}, 759 (1987).


\bibitem{Popov1991} V.V. Popov and O.V. Vasil'ev, Europhys. Lett. {\bf 15}, 7 (1991).

\bibitem{Woody1979} D.P. Woody and P.L. Richards, Phys. Rev. Lett. {\bf 42}, 925 (1979).

\bibitem{Glashow1983} H. Georgi, P. Ginsparg and S.L. Glashow, Nature {\bf 306}, 765 (1983); M. Axenides and R. Brandenberger, Phys. Lett. B {\bf 134}, 405 (1984).

\bibitem{Okun1982} L. B. Okun, Zh. Eksp. Teor. Fiz. {\bf 83}, 892 (1982)
[Sov. Phys. JETP {\bf 56}, 502 (1982)].

\bibitem{COBERocket1990} J.C. Mather {\it et al.}, Astrophys. J. Lett. {\bf 354}, L37 (1990); H.P. Gush, M. Halpern and E.H. Wishnow, Phys. Rev. Lett. {\bf 65}, 537 (1990).

\bibitem{StringReviews} S. Abel and J. Santiago, J. Phys. G {\bf 30}, R83 (2004); R. Blumenhagen {\it et al.}, Phys. Rep. {\bf 445}, 1 (2007).

\bibitem{Ahlers2007} M. Ahlers {\it et al.}, Phys. Rev. D {\bf 76}, 115005 (2007).

\bibitem{Jaeckel2008} J.~Jaeckel and A. Ringwald, Phys. Lett. B {\bf 659}, 509 (2008).

\bibitem{Ahlers2008} M. Ahlers {\it et al.}, Phys. Rev. D {\bf{77}}, 095001 (2008).

\bibitem{Bernardis1984} P. De Bernardis {\it et al.}, Astrophys. J. {\bf 284}, L21 (1984).

\bibitem{Angelis2002} A. De Angelis and R. Pain, Mod. Phys. Lett. A {\bf 17}, 2491 (2002).


\bibitem{Zavattini2007} E. Zavattini {\it et al.}, Phys. Rev. Lett. {\bf{96}}, 110406 (2006); {\it ibid.} {\bf 99}, 129901 (2007).

\bibitem{Robilliard2007} C. Robilliard {\it et al.}, Phys. Rev. Lett. {\bf 99}, 190403 (2007).

\bibitem{Raffelt1988} G. Raffelt and L. Stodolsky, Phys. Rev. D {\bf 37}, 1237 (1988).

\bibitem{Adler2008} S. L. Adler {\it et al.},
arXiv:0801.4739v4 [hep-ph], Ann. of Phys., in press (2008).


\bibitem{webLULI} See http://www.luli.polytechnique.fr/pages/LULI2000.htm

\bibitem{DeformableMirror} J.-P. Zou {\it et al.}, Appl. Opt. {\bf{47}}, 704 (2008).

\bibitem{Frings2006} P. Frings {\it et al.}, Rev. of Sc. Inst. {\bf 77}, 063903 (2006).

\bibitem{Batut2008} S. Batut {\it et al.}, IEEE Trans. Applied Superconductivity, {\bf 18}, 600 (2008).

\bibitem{SinglePhotonDetector} D. S. Bethune, W. P. Risk and G. W.
Pabst, J. Mod. Opt. {\bf{51}}, 1359 (2004).


\bibitem{FermiLab2008} A. S. Chou {\it et al.}, Phys.
Rev. Lett. {\bf 100}, 080402 (2008).

\bibitem{PVLAS2008} E. Zavattini {\it et al.}, Phys. Rev. D
{\bf 77}, 032006 (2008).

\bibitem{CavityRBF} S. DePanfilis {\it et al.}, Phys. Rev. Lett.
{\bf{59}}, 839 (1987); W. U. Wuensch {\it et al.}, Phys. Rev.
D {\bf{40}}, 3153 (1989).

\bibitem{CavityUF} C. Hagmann {\it et al.},
Phys. Rev. D {\bf{42}}, 1297 (1990).

\bibitem{KSVZ} J. E. Kim, Phys. Rev. Lett. {\bf 43}, 103 (1979); M.A.~Shifman, A.I.~Vainshtein and V.I.~Zakharov, Nucl. Phys. B {\bf 166}, 493 (1980).

\bibitem{DFSZ} M. Dine, W. Fischler and M. Srednicki, Phys. Lett. B {\bf 104}, 199 (1981); A.P.~Zhitnitskii, Sov. J. Nucl. Phys. {\bf 31}, 260 (1980).

\bibitem{CoulombLaw} G. D. Cochran and P. A. Franken,
Bull. Am. Phys. Soc. {\bf{13}}, 1379 (1968); D. F. Bartlett, P. E.
Goldhagen and E. A. Phillips, Phys. Rev. D {\bf{2}}, 483 (1970); E.
R. Williams, J. E. Faller and H. A Hill, Phys. Rev. Lett. {\bf{26}},
721 (1971).

\bibitem{Rydberg} R. G. Beausoleil {\it et al.}, Phys. Rev. A
{\bf{35}}, 4878 (1987).

\bibitem{Review_ParaExp} D. F. Bartlett and S. Logl, Phys. Rev.
Lett. {\bf{61}}, 2285 (1988).

\bibitem{BMVprototype} R. Battesti, {\it et al.}, Eur. Phys. J. D {\bf 46},
323 (2008).

\bibitem{QEDBMV} H. Euler and K. Kochel, Naturwiss. {\bf 23}, 246 (1935); W. Heisenberg and H. Euler, Z. Phys. {\bf 98}, 714 (1936); Z. Bialynicka-Birula and I. Bialynicka-Birula, Phys. Rev. D {\bf 2}, 2341 (1970); S.L. Adler, Ann. Phys. (N.Y.) {\bf 67}, 599 (1971).

\bibitem{Chameleon} P. Brax {\it et al.}, Phys. Rev. D {\bf 76}, 085010 (2007); M. Ahlers {\it et al.}, Phys. Rev. D {\bf 77}, 015018 (2008); H. Gies {\it et al.}, Phys. Rev. D {\bf 77}, 025016 (2008).

\bibitem{ELIwebsite} http://www.extreme-light-infrastructure.eu


\end{thebibliography}

\end{document}